\input harvmac.tex\overfullrule=0pt
\input labeldefs.tmp
\writedefs
\def\d{{\rm d}}
\def\Im{\mathop{\rm Im}\nolimits}
\def\e#1{{\rm e}^{#1}}
\def\E#1{{\rm e}^{\textstyle #1}}
%
%
\input epsf
\long\def\ffig#1#2#3{%
\xdef#1{\the\figno}%
\writedef{#1\leftbracket \the\figno}%
\midinsert%
\parindent=0pt\leftskip=1cm\rightskip=1cm\baselineskip=11pt%
\centerline{#3}
\vskip 8pt\ninepoint%
{\bf Fig.\ \the\figno:} #2%
\endinsert%
\goodbreak%
\global\advance\figno by1%
}
\long\def\fig#1#2#3{\ffig{#1}{#2}{\epsfbox{#3}}}
%
\def\pre#1{ (preprint {\tt #1})}
\lref\BIPZ{E.~Br{\'e}zin, C.~Itzykson, G.~Parisi and J.-B.~Zuber,
{\it Commun. Math. Phys.} 59 (1978), 35.}
\lref\Kaz{V.A.~Kazakov, {\it Phys. Lett.} A119 (1986), 140.}
\lref\ZUB{J.-B.~Zuber in Les Houches 1988 proceedings, Session XLIX,
Editors E.~Br{\'e}zin and J.~Zinn-Justin, North-Holland.}
\lref\DAL{S.~Dalley, {\it Mod. Phys. Lett.} A7 (1992), 1651.}
\lref\GK{D.~Gross and I.~Klebanov, {\it Nucl. Phys.} B344 (1990), 475.}
\lref\KAZMIG{V.A.~Kazakov and A.~Migdal, {\it Nucl. Phys.} B311 (1988), 171.}
\lref\KoS{I.K.~Kostov, {\it Mod. Phys. Lett.} A4 (1989), 217\semi
M.~Gaudin and I.K.~Kostov, {\it Phys. Lett.} B220 (1989), 200\semi
I.K.~Kostov and M.~Staudacher, {\it Nucl. Phys.} B384 (1992), 459.}
\lref\KP{V.A.~Kazakov and P.~Zinn-Justin, {\it Nucl. Phys.} B546 (1999),
647\pre{hep-th/9808043}.}
\lref\PZJZ{P.~Zinn-Justin and J.-B.~Zuber, preprint {\tt math-ph/9904019},
to appear in the proceedings of the 11th 
International Conference on Formal Power Series and Algebraic 
Combinatorics, Barcelona June 1999\semi
P.~Zinn-Justin, preprint {\tt math-ph/9910010}, to appear
in the proceedings of the MSRI semester on Random Matrices 1999.}
\lref\KKN{V.A. Kazakov, I.K. Kostov and N. Nekrasov,
preprint {\tt hep-th/9810035}.}
\lref\EYK{B. Eynard and C. Kristjansen,
{\it Nucl. Phys.} B516 (1998), 529.}
\lref\AND{N. Andrei, K. Furuya and J.H. Lowenstein,
{\it Rev. Mod. Phys.} 55 (1983), 331.}
\lref\YY{C.N. Yang and C.P. Yang, {\it Phys. Rev.} 150 (1966), 327.}
\lref\Kazb{V.A.~Kazakov, {\it Nucl. Phys.} B (Proc. Suppl.) 4 (1988), 93.}
\lref\PZJ{P. Zinn-Justin, preprint {\tt cond-mat/9903385}, to appear
in {\it J. Stat. Phys.}}
\lref\KoKa{I.K.~Kostov, unpublished notes\semi
V.A.~Kazakov and I.K.~Kostov, work in progress.}
\lref\EZJK{B. Eynard and J. Zinn-Justin, 
{\it Nucl. Phys.} B386 (1992), 558\pre{hep-th/9204082};
B. Eynard and C. Kristjansen,
{\it Nucl. Phys.} B455 (1995), 577\pre{hep-th/9506193}.}
\lref\DAUL{J.M.~Daul, preprint {\tt hep-th/9502014}.}
\lref\DAV{F. David, {\it Nucl. Phys.} B257 [FS14] (1985), 45.}
\lref\KAZa{V.A. Kazakov, {\it Phys. Lett.} B150 (1995), 282.}
\Title{\vbox{\hbox{RUNHETC-99-36}\hbox{{\tt cond-mat/9909250}}}}
{{\vbox {
\vskip-10mm
\centerline{The six-vertex model on random lattices}
}}}
\medskip
\centerline{P.~Zinn-Justin }\medskip
\centerline{\it Department of Physics and Astronomy, Rutgers University,} 
\centerline{\it Piscataway, NJ 08854-8019, USA}
\bigskip
\noindent
In this letter, the 6-vertex model on dynamical random lattices is
defined via a matrix model and rewritten (following I.~Kostov)
as a deformation of the
$O(2)$ model. In the large $N$ planar limit, an exact solution is found
at criticality. The critical exponents of the model are determined;
they vary continously along the critical line.
The vicinity of the latter is explored, which confirms that
we have a line of $c=1$ conformal field theories coupled to gravity.

\Date{September 1999}\def\rem#1{}

\newsec{Introduction}
Starting from the observation that the combinatorial
properties of large $N$ matrix models \BIPZ\ allow them
to reproduce summations over discretized surfaces \refs{\DAV,\KAZa}
it is possible
to study various 2D statistical models on random dynamical lattices,
that is consider systems in which both the ``spin'' degrees of freedom
sitting on the lattice and the lattice itself are allowed to fluctuate.
Among the models that were solved this way, let us cite
the Ising model \Kaz, the $O(n)$ model \refs{\KoS,\EZJK}, the Potts
model \refs{\Kazb,\DAUL,\PZJ}.
All these models have critical points which correspond to conformal
field theories of central charge $c$ less or equal to $1$ coupled
to gravity, the limiting case $c=1$ being of particular interest.
One well-known statistical
model has until now resisted attempts at an exact solution
(on random lattices): the 8-vertex
model and its critical version, the 6-vertex model. On a flat lattice,
the 6-vertex model displays an infra-red behavior which spans
a whole semi-infinite line of $c=1$ theories; a similar behavior
is expected when put on random lattices \DAL.
In fact, a special
2-parameter slice of the 8-vertex model was solved exactly
in \KP; the 6-vertex point of the model was shown to indeed exhibit
a $c=1$ behavior.

In this letter, we shall perform a study of a matrix model
which describes the 6-vertex model. We shall be
concerned with its large $N$ limit,
which correspond to selecting the spherical topology for the lattices.
We shall solve the model exactly when the
renormalized cosmological constant vanishes, that is in the limit
where the average size of the graphs goes to infinity, and
give explicit expressions for some averages. We shall then
show how to explore the vicinity of this critical region
and compute the first correction, which yields the string suceptibility.

\newsec{The model and its saddle point equation}
The model is defined by the following partition function:
\eqn\defZ{Z=\int \d X \d X^\dagger \exp\left[N\tr\left(
-XX^\dagger + b X^2X^\dagger{}^2+{c\over 2}(XX^\dagger)^2\right)\right]}
where $X$ is a general $N\times N$ complex matrix.
The Feynman rules of the model reproduce the configurations
of the {\it six-vertex model} (figure \six) on
a random four-valent lattice.
\epsfxsize=9cm\fig\six{Part of a diagram generated
by the perturbative expansion of \defZ.}{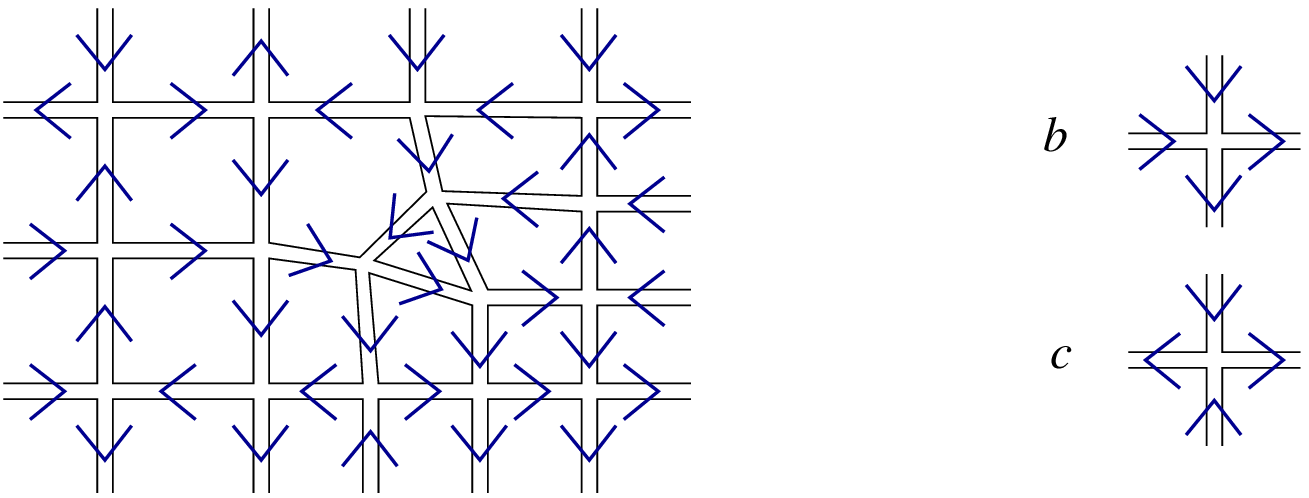}

In the parametrization of e.g.\ \ZUB,
$c/b=2\cos(\lambda\pi/2)$; the parameter $\lambda$ can,
on random surfaces, vary in the range $0\le\lambda<2$.
For $\lambda=0$ we recover the usual $O(2)$ model \KoS;\rem{Note that
in this formulation of the $O(2)$ model,
since propagators have arrows,
the double counting of each loop is accounted for
by the two possible orientations of the
loops.}
for $\lambda=1$ we recover the critical $ABAB$ model \KP.\rem{The previous
remark also applies to $\lambda=1$ except this time
loops {\it intersect}\/ each other at each vertex.}
Finally, the singular limit $\lambda\to 2$ corresponds to the
three-colouring problem \refs{\EYK,\KKN}.
When $\lambda$ is fixed, we can still vary $b$, which plays the role
of (bare) cosmological constant. For any $\lambda$ we expect that
there is a value $b_{\rm crit}(\lambda)$ for which the average size of the
graphs diverges.

In order to solve the model, we use the following trick:
we decouple the quartic interaction \KoKa
\eqn\dec{Z=\int \d A \d X\d X^\dagger \exp\left[N\tr\left(
-XX^\dagger-{1\over 2}A^2+\sqrt{b} A(XX^\dagger \e{i\lambda\pi/4}+
X^\dagger X \e{-i\lambda\pi/4})\right)\right]}
where $A$ is a hermitean matrix. This new model
is a deformation of the $O(2)$ model. Its Feynman rules (figure \lps)
allow to interpret it as a model of oriented loops in which each
left/right turn costs $\omega^{\pm 1}
=\e{\pm i\lambda\pi/4}$; this means that each loop,
taking into account its two possible orientations, contributes
a factor of
$2\cos(p\lambda\pi/4)$, where $p$ is the number of right turns minus
the number of left turns performed when going around the loop.
For a regular infinite lattice, this number
is fixed and we recover this way the usual $O(n)$ model with 
$n=2\cos(p\lambda\pi/4)$; but this connection
between the 6-vertex and the $O(n)$ models
breaks down on a random lattice (because of curvature). It only
remains at $\lambda=0$, of course, where we recover, as mentioned previously,
the $O(2)$ model.
\epsfxsize=7.5cm\fig\lps{Part of a diagram generated
by the perturbative expansion of \dec.}{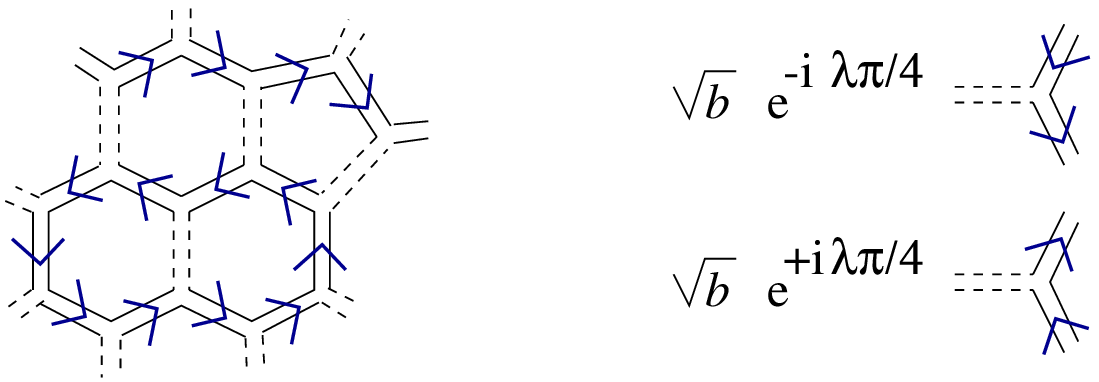}

The next step to solve the model is to integrate out $X$ and $X^\dagger$,
and shift $A$ by the constant $\gamma={1\over\sqrt{b}(\omega+\omega^{-1})}$:
\eqn\ZA{Z=\int \d A \det{}^{-1}\left(\omega\otimes A+
A\otimes\omega^{-1}\right)
\exp\left(-{N\over2}\tr (A-\gamma)^2\right)
}
In terms of the eigenvalues $a_i$ of $A$:
\eqn\eig{Z=\int
\prod_{i=1}^N \d a_i \E{-{N\over2}(a_i-\gamma)^2}
{\prod_{i\ne j} (a_i-a_j)
\over\prod_{i,j} \left(\omega a_i+\omega^{-1}a_j\right)}
}

In the large $N$ limit, the $a_i$ form a saddle point distribution
characterized by a continuous density $\rho_0(a)\d a$
which fills an interval $[\alpha,\beta]$ of the real line. 
It will be convenient
to make the following change of variable: $a=\beta\, \e{u}$.
Up to an overall constant we find:
\eqn\shift{Z=\int
\prod_{i=1}^N \d u_i \E{-{N\over 2}(\beta\, \e{u_i}-\gamma)^2}
\prod_{i\ne j} {\e{u_i}-\e{u_j}
\over \omega \e{u_i}+\omega^{-1}\e{u_j}}
}
One notices that the two-body interaction now only
depends on the difference $u_i-u_j$.
More explicitly, we are now trying to minimize an action
of the form
\eqn\acti{S=\int\d u\rho(u)\,V(u)
+{1\over2}\int\hskip-0.2cm\int\d u\rho(u)\d v\rho(v)\,
V_2(u-v)}
with the density $\d u \rho(u)=\d a \rho_0(a)$,
a potential $V(u)={1\over2}(\beta\,\e{u}-\gamma)^2$ and an interaction
\eqn\inter{
V_2(u)=-2\log(1-\e{u})+\log(1+\omega^2\e{u})+\log(1+\omega^{-2}\e{u})
}
The density $\rho(u)$ satisfies $\rho(u)\ge 0$ and $\int\d u\rho(u)=1$.
It is worth remarking the similarity of this problem with the determination
of the ground state in the presence of magnetic field
in Bethe Ansatz solvable models. We shall comment on this analogy later.

On the support of $\rho(u)$, 
the minimization of $S$ leads to the saddle point
equation:
\eqn\spe{K\star\rho(u)=\beta\,\e{u}(\beta\,\e{u}-\gamma)}
where $\star$ means convolution product and $K$ is the derivative of $-V_2$:
\eqn\kernel{K(u)={2\over 1-\e{-u}}-{1\over 1+\omega^2\e{-u}}
-{1\over 1+\omega^{-2}\e{-u}}
}
Principal part at $u=0$ is implied.

We shall now proceed to solve this equation.

\newsec{Exact results at criticality}
The analytic problem \spe\ is well known to be exactly
solvable using the Wiener--Hopf technique
when the support of
$\rho(u)$ is semi-infinite. In our case this corresponds
to $\alpha=0$, which is precisely the critial regime (i.e.\ when
the renormalized cosmological constant vanishes and the area
of the lattices becomes large). The support
of $\rho(u)$ is then $[-\infty,0]$ (figure \well). Note that the situation
is then analogous to the Bethe Ansatz study of the ground state
in the presence of magnetic field of 2D integrable models
with only one chirality, see for example \AND.
\epsfxsize=4cm
\fig\well{Potential $V(u)$. The eigenvalues fill the well; criticality
is attained when the eigenvalues start ``overflowing'' at minus infinity.}
{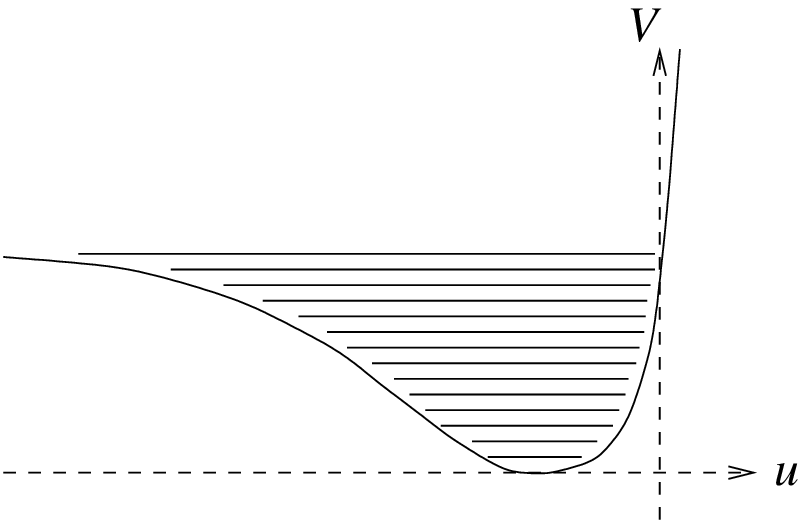}

We introduce the Fourier transform $\hat{K}(k)$:
\eqn\four{\eqalign{
\hat{K}(k)&=\int_{-\infty}^{+\infty}K(u)\e{iku}\d u\cr
&=4\pi {\sinh({1\over2}+{\lambda\over4})\pi k
\sinh({1\over2}-{\lambda\over4})\pi k
\over\sinh\pi k}\cr}
}
Similarly we have the Fourier transform $\hat{\rho}(k)$
of $\rho(u)$, which
can be defined alternatively as an average in our model:
\eqn\avg{
\hat{\rho}(k)={1\over\beta^{ik}}\left<{1\over N} 
\tr A^{ik} \right>
}
It is clear that $\hat{\rho}(k)$ is an analytic function
in the lower half-plane $\Im k\le 0$. We now Fourier transform the 
equation \spe:
\eqn\fourspe{
\hat{K}(k)\hat{\rho}(k)=\hat{f}(k)
}
where $\hat{f}(k)$ is a function whose inverse
Fourier transform $f(u)$ satisfies
$f(u)=\beta\e{u}(\beta\,\e{u}-\gamma)$ for $u\le 0$.

In order to determine $\hat{\rho}$ and $\hat{f}$, we decompose
$\hat{K}(k)$ as $\hat{K}(k)=\hat{K}_-(k)/\hat{K}_+(k)$
where $\hat{K}_-$ (resp.\ $\hat{K}_+$) 
is holomorphic in the lower (resp.\ upper)
half-plane. Explicitly, we choose
\eqn\Kpm{\eqalign{
\hat{K}_-(k)=i(k-i0)
{\Gamma(1+ik)\over \Gamma(1+i u_+ k)\Gamma(1+i u_- k)}
\e{i\epsilon k}&\cr
\hat{K}_+(k)={4\over\pi^2}{1\over 4-\lambda^2}
{\Gamma(1-iu_+ k)\Gamma(1-iu_- k)\over
\Gamma(1-ik)} 
\e{i\epsilon k}&\cr
}
}
where $u_\pm ={1\over 2}\pm {\lambda\over4}$ and
$\epsilon=u_+\log u_+ + u_- \log u_-$.
From \fourspe\ we infer
$$\hat{K}_-(k)\hat{\rho}(k)=\hat{K}_+(k)\hat{f}(k)$$
The left hand side is a function which is holomorphic in the
lower half-plane, and therefore the right hand side must also be.
It is now an easy exercise to find it starting from
$f(u)=\beta(\beta\,\e{u}-\gamma)$, $u\le 0$. The result is:
\eqn\solrhs{
\hat{K}_-(k)\hat{\rho}(k)=\beta\left(\beta {K_2\over i(k-2i)}
-\gamma{K_1\over i(k-i)}\right)
}
where $K_1$ and $K_2$ are $\hat{K}_+(k)$ evaluated at
$k=i$ and $k=2i$:
$K_1={1\over 4\pi\cos(\lambda\pi/4)} \e{-\epsilon}$
and $K_2=
{\lambda\over 4\pi\sin(\lambda\pi/2)} \e{-2\epsilon}$.
Dividing by $\hat{K}_-(k)$ gives $\hat{\rho}(k)$.
Note however that one must impose the
normalization condition $\hat{\rho}(k=0)=1$. This in fact imposes
two constraints since generically the function $\hat{\rho}(k)$ given by
\solrhs\ has a pole at $k=0$. The pole cancellation condition reads
\eqn\condone{
\gamma=\beta{K_2\over 2 K_1}
}
while the normalization condition reads
\eqn\condtwo{
\hat{\rho}(0)={1\over4}
\beta^2 K_2=1
}
These two conditions determine the critical values of $\beta$ and $\gamma$.
In terms of the original coupling constant $b$, we find:
\eqn\bcrit{
b_{\rm crit}={1\over 32}{\sin(\lambda\pi/4)\over\lambda\pi/4}
{1\over\cos^3(\lambda\pi/4)}
}
Finally we obtain the expression for $\hat{\rho}(k)$:
\eqn\solrho{
\hat{\rho}(k)=2
{\Gamma(1+iu_+ k)\Gamma(1+iu_- k)\over
\Gamma(3+ik)} 
\e{-i\epsilon k}
}

The usual correlation functions $\left<{1\over N}\tr A^n\right>$ are simply
given, according to \avg, by $\beta^n \hat{\rho}(k=-in)$. However,
to extract the universal information from the correlation functions,
it is simpler to consider the singularities of
the density.

The density $\rho_0(a)$ has two singularities: one at $a=\alpha=0$ and
one at $a=\beta$, which correspond to $u=-\infty$ and $u=0$.

The singularity of $\rho_0(a)$ at $a=\beta$ is
determined by the behavior of $\hat{\rho}(k)$ as $k\to\infty$. We find
that $\hat{\rho}(k)\buildrel k\to\infty\over\sim k^{-3/2}$, or
\eqn\sqrsing{
\rho_0(a)\sim (a-\beta)^{1/2}
}
i.e.\ the usual square root singularity.

On the other hand, the singularity of $\rho_0(a)$ at $a=0$ can
be inferred from the poles of $\hat{\rho}(k)$, which lie at
\eqn\poles{
k={i\, n\over u_\pm } \qquad n=1,2,\ldots
}
For $\lambda>0$, the pole closest to the real axis is
$k=i/u_+$, which corresponds to a behavior
$\rho(u)\buildrel u\to-\infty\over \sim
\e{u/u_+}$
and therefore
$\rho_0(a)\sim a^{{1\over u_+}-1}$ or more explicitly
a leading singularity of the form
\eqn\sing{
\rho_0(a)\sim a^{\scriptstyle 2-\lambda\over\scriptstyle 2+\lambda}
}
Therefore we have found one critical exponent of our model,
which depends continuously on $\lambda$.
The other poles give subleading terms in the expansion around
$a=0$. 

In the limiting case $\lambda=0$, we find a double pole at $k=2i$,
which results in a behavior $\rho(u)\sim u\,\e{2u}$, or
\eqn\singlim{
\rho_0(a)\sim a\log a
}

\newsec{Vicinity of the critical line}
Outside the critical regime, i.e.\ when the support $[\alpha,\beta]$
of the density $\rho(u)$ is finite, it is not clear how to solve exactly
the equation \spe. This is completely analogous to the Bethe Ansatz
equations for 2D integrable models with 2 chiralities (like massive
relativistic models). However, what one can do is an exact expansion
as $\alpha\to 0$. Here we shall follow the method of \YY.
We shall only compute the first correction to the calculation of
the previous section, up to some constants which could
be determined by a more careful study.

Let us denote $B=\log(\beta/\alpha)$. We are interested in the
limit where $B$ is large (in the Bethe Ansatz language, the limit
where the two chiralities are almost decoupled from each other). Then
we know that we can split our equation \spe\ into two equations
for the regions $u\approx -B$ and $u\approx 0$. The first equation
is after the shift $u\to u+B$:
\eqn\spea{
K\star\rho(u)=-\beta\gamma\e{-B} \e{u}\qquad \forall u\in[0,B]
}
We introduce the function $g(u)$ with support $[-\infty,0]$ such that
\eqn\speaa{
K\star\rho(u)=-\beta\gamma\e{-B} \e{u}+\e{-B/u_+} g(u)
\qquad\forall u\in[-\infty,B]}
We have included the prefactor $\e{-B/u_+}$ so that
$g(u)$ has a finite limit when $B\to\infty$, as a simple calculation shows
(for $\lambda=0$ one should replace $\e{-B/u_+}$ with $B\e{-2B}$).
The second equation is our usual saddle point equation; we
can insert in it the first correction:
\eqn\specor{
K\star\rho(u)=
\beta\,\e{u}(\beta\,\e{u}-\gamma)
+\e{-B/u_+} g(u+B)\qquad \forall u\in[-\infty,0]
}
so that the equation is valid
for all negative $u$. Therefore this is again a Wiener--Hopf problem.
We shall not bother to solve it explicitly; let us simply
write down the form of the solution at leading order as $B\to\infty$:
\eqn\specorb{
\hat{K}_-(k)\hat{\rho}(k)=\beta\left(\beta {K_2\over i(k-2i)}
-\gamma{K_1\over i(k-i)}\right)+\e{-B/u_+}\e{-ikB} \hat{h}(k)
}
When dividing by $\hat{K}_-(k)$ we are again faced with the problem
of the behavior at $k=0$, which leads to the two conditions:
\eqn\twocond{
\eqalign{
\gamma&=\beta{K_2\over 2 K_1}+c_1 \e{-B/u_+}+\cdots\cr
1&={1\over4}
\beta^2 K_2(1+c_2 B\e{-B/u_+}+c_3 \e{-B/u_+}+\cdots)\cr
}}
where $c_1$, $c_2$, $c_3$ are constants.
The term $B\e{-B/u_+}$ comes from differentiation of $\e{-ikB}$. This allows
to solve for $\beta$ and $\gamma$ as a function of $B$. The renormalized
cosmological constant $\Delta$ is defined as the variation of $\gamma$.
From \twocond\ we infer
\eqn\cosm{\Delta\equiv \gamma-\gamma_{\rm crit}\sim B\e{-B/u_+}}
Next, we consider any correlation function 
$\left<{1\over N}\tr A^n\right>$
(obtained by evaluating $\hat{\rho}(k)$). It is clear from
Eq.~\specorb\ that its variation is of the form
\eqn\var{
\hat{\rho}(k)= c_4 + c_5 B\e{-B/u_+}+ c_6 \e{-B/u_+}+\cdots}
and therefore its singular part is
\eqn\varb{
\hat{\rho}(k)_{\rm sing} \sim
{\Delta\over \log\Delta}
}
This last result displays a zero string susceptibility exponent
and a logarithmic correction characteristic of a $c=1$ theory \KAZMIG.

\newsec{Summary of results and prospects}
We have given in this letter the exact solution of the six
vertex model on random planar graphs at zero cosmological constant
and described the vicinity of this critical line. 
Note that all the expressions found agree with the known results
in the $\lambda=0$ and $\lambda=1$ cases. In particular, for $\lambda=0$
the critical value $b_{\rm crit}={1\over32}$ and the behaviors \singlim\ 
coincide with earlier calculations \KoS, while for $\lambda=1$
the critical value $b_{\rm crit}={1\over4\pi}$ and some correlation functions
such as $\left<{1\over N}\tr XX^\dagger\right>={\pi\over 2}(4-\pi)$
coincide with \KP. 
The limit $\lambda\to 2$ is singular, as can be seen in \bcrit,
and therefore we cannot compare directly our results with those
of \refs{\EYK,\KKN};
in fact, even though the $\lambda=2$ probably {\it does} have
a critical point for a finite value $b_{\rm crit}$ \KKN, it is believed
to belong to the universality class of pure gravity, which is
different from
the critical behavior found for $\lambda<2$ ($c=1$ theories).

The solution is explicit enough to allow us to give exact expressions
for correlation functions at criticality, and the method can be used
to generate an exact expansion around the critical region.
Two critical exponents have been computed this way: the exponent
governing the singularity of the density of eigenvalues, and
the string susceptibility exponent; the latter turned out to be
zero plus a logarithmic scaling violation, confirming the central charge
$c=1$ of the infra-red CFT.

This raises the hope that it is possible to solve much more general models.
For example, one can replace the quadratic potential
of $A$ with a more general one. This could be used to simulate e.g.\ vortices
in the 6-vertex model or dilution in the deformed $O(2)$ model.
There is in principle no problem to finding the critical properties of
these models, and we expect a rich structure of multi-critical points.
All this should appear in a future publication, as well as a more
detailed description of the off-critical region.

Finally, let us mention that these results should have an interesting
application to knot theory and more precisely the counting
of alternating links, see \PZJZ. This is currently under study.

\vskip1cm
\centerline{\bf Acknowledgements}
I would like to thank N.~Andrei for his advice and encouragement,
I.~Kostov for a careful reading of the draft,
and A.~Its, V.~Kazakov for discussions.
This work was supported in part by the DOE grant DE-FG02-96ER40559.

\listrefs
\bye